\documentclass[preprint,12pt]{elsarticle}
\usepackage{CJK,lineno}
\usepackage{amsmath}
\usepackage{amssymb}
\usepackage[pdftex,colorlinks]{hyperref}
\usepackage{multirow}

\newtheorem{Theo}{Theorem}

\newenvironment{Proof}{{\noindent \em Proof}}{\hfill $\Box$\par}
\biboptions{compress}
\begin{document}
\begin{frontmatter}
\title{Quantum algorithms for the Goldreich-Levin learning problem}
\author{Hongwei Li}
\address{School of Mathematics and Statistics, Henan Finance University, Zhengzhou, 450046, Henan, China}
\begin{abstract}
The Goldreich-Levin algorithm was originally proposed for a cryptographic purpose and then applied to learning. The algorithm is to find some larger Walsh coefficients of an $n$ variable Boolean function. Roughly speaking, it takes a $poly(n,\frac{1}{\epsilon}\log\frac{1}{\delta})$ time to output the vectors $w$ with Walsh coefficients $S(w)\geq\epsilon$ with probability at least $1-\delta$. However, in this paper, a quantum algorithm for this problem is given with query complexity $O(\frac{\log\frac{1}{\delta}}{\epsilon^4})$, which is independent of $n$.
Furthermore, the quantum algorithm is generalized to apply for an $n$ variable $m$ output Boolean function $F$ with query complexity $O(2^m\frac{\log\frac{1}{\delta}}{\epsilon^4})$.
\end{abstract}


\end{frontmatter}
\section{Introduction}
\vspace*{-0.5pt}
\noindent
It was Deutsch \cite{Deu85} who first gave a quantum algorithm, which demonstrated a quantum computer could compute faster than a classical computer. Later, the algorithm was improved to Deutsch--Jozsa algorithm \cite{DJ92,CEMM98} and some other quantum algorithms were proposed for different problems. Bernstein and Vazirani \cite{BV93} gave a quantum algorithm for learning the expression of a Boolean function $f=a\cdot x$ with one query to the oracle using the same circuit as the Deutsch--Jozsa algorithm did, while the classical algorithm should query $O(n)$ times and then solve an linear equation. The Deutsch--Jozsa algorithm shows an exponential speedup over the best known classical algorithm, and the Bernstein--Vazirani algorithm performs a polynomial speedup than classical algorithm. Here we investigate how the Bernstein--Vazirani algorithm could be generalized to work on a function $f$ correlated with multiple linear functions, this is the Goldreich--Levin learning algorithm \cite{O14}.

The Goldreich--Levin problem \cite{Bell99} was originally presented for the purpose of cryptography. Roughly speaking, the task of the algorithm is to determine a string $a$ by querying two given oracles IP (inner product) and EQ (equivalance) about $a$. \cite{Gol99} showed an algorithm with $O(n/\epsilon^2)$ IP queries and $O(1/\epsilon^2)$ EQ queries. \cite{AC02} proposed a quantum algorithm to solve the Goldreich-Levin problem with $O(1/\epsilon)$ IP queries and $O(1/\epsilon)$ EQ queries. \cite{ACIP06} proved the above algorithm was optimal.

Later, the Goldreich--Levin algorithm developed for finding large Walsh coefficients of a Boolean function \cite{O14}. Boolean functions are widely used in symmetric cryptography and error  correcting codes. Almost all the properties of Boolean functions can be connected to Walsh spectra of Boolean functions. Classically, there is a divide-and-conquer butterfly algorithm (so called Fast Wash Transform) \cite{Car08} to compute the Walsh spectrum of an $n$ variable Boolean function $f$ with time complexity $n2^n$. Usually, the large Walsh coefficients of a Boolean function play an important role in the properties of Boolean functions. The Goldreich--Levin probabilistic algorithm \cite{O14} outputs some large Walsh coefficients of $f$ in time $poly(n,\frac{1}{\epsilon}\log\frac{1}{\delta})$. Here, we investigate a quantum algorithm about this problem for a Boolean function then generalize it to a multi-output Boolean function.

There have been quantum algorithms for the same large coefficients finding problem of a (multi-output) Boolean function $f$ in \cite{MO10,CGXL19}, but the methods they used were quite different from here in this paper, they all used a divide and conquer strategy similarly to the classical algorithm in \cite{O14}. The difference between these quantum algorithms and the classical algorithm was the way used to estimate the Walsh coefficients $S_f(x)$, using quantum circuits contrast to classical circuits. The classical algorithm \cite{O14} then obtained a sample from the distribution related to $f(x)$, and the quantum algorithms \cite{MO10,CGXL19} using a Grover-like operator to amplify the amplitude to get an estimation of $S_f(x)$.

Hillery and Andersson \cite{HA11} gave a quantum algorithm to test weather a Boolean function $f$ is linear or not using the Bernstein--Vazirani algorithm combined with Grover operator. This is a quantum algorithm worked on possible non-linear functions.
Inspired partly by that, a quantum algorithm about large coefficients finding problem is proposed with time entirely unrelated to $n$. Then, the algorithm is fine-tuned to work on multi-output Boolean functions.
\section{Preliminaries}
\subsection{Notations and Definitions}
\noindent
Let $n$ be a positive integer. $F_2$ is a finite field with two elements $\{0,\,1\}$, and $F_2^n$ is a vector space over $F_2$. A mapping $f:F_2^n\rightarrow F_2$ is usually called a Boolean function.

For
$a=(a_1,\,\ldots,\,a_n),\,b=(b_1,\,\ldots,\,b_n)$, define
$a\cdot b=a_1b_1\oplus\cdots\oplus a_nb_n$
as the inner product of $a$ and $b$, where $\oplus$ is sum module 2.

Define
\begin{equation}\label{eq:a}
S_f(a)=2^{-n}\sum_{x\in F_2^n}(-1)^{a\cdot x\oplus f(x)}
\end{equation}
as the Walsh transform of $f$.

\subsection{Deutsch--Jozsa algorithm \cite{DJ92}}
Suppose $f$ is a Boolean function, and it is balanced or constant. Deutsch-Jozsa algorithm is to decide which one is the case through only one measurement. The specific steps of the algorithm are as follows.

1. Perform the Hadamard transform $H^{(n+1)}$ on  $|\psi_0\rangle=|0\rangle^{\otimes n}|1\rangle$ to obtain
\begin{equation}\label{eq:b}|\psi_1\rangle=\sum_{x\in F^n_2}
\frac{|x\rangle}{\sqrt{2^n}}\cdot\frac{|0\rangle-|1\rangle}{\sqrt{2}}.\end{equation}

2. Apply the $f$-controlled-NOT gate on $|\psi_1\rangle$ producing
\begin{equation}\label{eq:c}
|\psi_2\rangle=\sum_{x\in F^n_2}
\frac{(-1)^{f(x)}|x\rangle}{\sqrt{2^n}}\cdot\frac{|0\rangle-|1\rangle}{\sqrt{2}}.
\end{equation}

3. Apply $n$ Hadamard gates to the first $n$ qubits to get
\begin{equation}\label{eq:d}
\begin{split}
|\psi_3\rangle &=\sum_{w\in F^n_2}\frac{1}{2^n}\sum_{x\in F^n_2}(-1)^{f(x)+w\cdot x}|w\rangle
\cdot\frac{|0\rangle-|1\rangle}{\sqrt{2}}\\
&=\sum_{w\in F^n_2}S_f(w)|w\rangle
\cdot\frac{|0\rangle-|1\rangle}{\sqrt{2}}.
\end{split}
\end{equation}

4. Measure the first $n$ qubits of $|\psi_3\rangle$ in the computational basis.

If we get the zero state, the function is constant, otherwise it is balanced.

\noindent\textbf{Bernstein--Vazirani algorithm \cite{BV93}}

If the function $f$ given to the oracle has an expression $f=a\cdot x$, then
$S_f(a)=1$, $S_f(b)=0$ for any $b\neq a$ by equation \eqref{eq:a}. Therefore,
running the above algorithm will yield the vector $a$.

\noindent\textbf{Remark 1}\quad If there is no promise about $f$, the aforementioned procedures will result a vector $w$ with probability $S_f^2(w)$.

\noindent\textbf{Example 1}\quad If $f(x_1,x_2,x_3,x_4)=x_1+x_2+x_2x_3+x_3x_4$, then $S_f(1001)=S_f(1100)=S_f(1110)=\frac{1}{2}$, $S_f(1011)=-\frac{1}{2}$. Running the above algorithm will obtain $1001, 1100, 1110, 1011$ with probability $\frac{1}{4}$.
\section{Quantum Goldreich-Levin theorem}
In this section, we give quantum algorithms producing larger Walsh coefficients of an $n$ variable (multi-output) Boolean function $f$. The query complexity of the algorithm is independent with $n$, such an complexity has not been seen in the literature.
\subsection{Quantum Goldreich-Levin theorem for a Boolean function}
Now, based on Deutsch--Jozsa algorithm, we present our algorithm. In fact, we get a sample from the probability distribution $P$ with
$P(a)=S_f^2(a)$ for every $a\in F_2^n$.
\begin{bframe}
\textbf{Algorithm 1}

 For any $ 0<\epsilon\leq 1$, $0<\delta<1$,
let $l=\frac{8\log\frac{1}{\delta}}{\epsilon^4}$,
$s=\frac{\epsilon^2l}{2}=\frac{4\log\frac{1}{\delta}}{\epsilon^2}$.
$H=\emptyset$, $L=\emptyset$ are two sets, where $\emptyset$ is the empty set.

\quad for all $k\in [\,1,\,l\,]$ do

\quad\quad Run the Deutsch--Jozsa algorithm  to get $n$-bit vector $w$;

\quad \quad If $w\in H$, then

\quad\quad\quad $i_w=i_w+1$;

\quad \quad else

\quad\quad\quad update $H:=H\bigcup\{w\}$;

\quad\quad end if

\quad\quad If $i_w\geq s$, then

\quad\quad\quad update $L:=L\bigcup\{w\}$;

\quad\quad end if

\quad end for

\quad Output $L$.

\end{bframe}

\begin{Theo}\label{th:a}
Given a Boolean function $F:\{0,\,1\}^n\rightarrow\{0,\,1\}$ ,given $0<\epsilon\leq 1$, $0<\delta<1$, running the above Algorithm 1
output a list $L=\{w_1,\ldots,w_t\}$ such that
\begin{equation}\label{eq:e}
\begin{cases}
|S_f(w)|\geq \epsilon\Rightarrow w\in L,\\
w\in L\Rightarrow |S_f(w)|\geq \epsilon/2.
\end{cases}
\end{equation}
with probability at least $1-\delta$ and the query complexity is $O(\frac{\log\frac{1}{\delta}}{\epsilon^4})$.
\end{Theo}
\begin{Proof}
\qquad Each running of the Deutsch--Jozsa algorithm is a randomized trial.
For arbitrary fixed $w_0\in F_2^n$, $S_f^2(w_0)$ is the probability of obtaining $w_0$ through a trial. Let $X$ be a random variable defined below.
\begin{equation}\label{eq:f}
X=
\begin{cases}
1 & w=w_0,\\
0 & w\neq w_0.
\end{cases}
\end{equation}
Then the mathematical expectation of $X$ is $E(X)=S_f^2(w_0)$. To $l$ times running the Deutsch--Jozsa algorithm there correspond $l$ independent identically distributed random variables $X_1,\,X_2,\,\cdots X_l$.

If $|S_f^2(w_0)|\geq \epsilon$, then $E(X)=S_f^2(w_0)\geq \epsilon^2$. By the Hoeffding inequality \cite{W63}, we have
\begin{equation}\label{eq:g}
\mathrm{Pr}(S_f^2(w_0)-\frac{1}{l}\sum_{i=1}^lX_i<\frac{\epsilon^2}{4})\geq
1-e^{-2l(\frac{\epsilon^2}{4})^2},
\end{equation}
therefore,
\begin{equation}\label{eq:h}
\mathrm{Pr}(\frac{1}{l}\sum_{i=1}^lX_i>S_f^2(w_0)-\frac{\epsilon^2}{4}>
\frac{\epsilon^2}{2})\geq 1-\delta,
\end{equation}
i.e.,
\begin{equation}\label{eq:i}
\mathrm{Pr}(\sum_{i=1}^lX_i> \frac{l\epsilon^2}{2})\geq 1-\delta.
\end{equation}

On the other side, if $w_0\in L$, i.e., $\sum_{i=1}^lX_i \geq \frac{l\epsilon^2}{2}$, we can obtain the following similar result by the Hoeffding inequality
\begin{equation}\label{eq:j}
\mathrm{Pr}(\frac{1}{l}\sum_{i=1}^lX_i-S_f^2(w_0)<\frac{\epsilon^2}{4})\geq
1-e^{-2l(\frac{\epsilon^2}{4})^2},
\end{equation}
This is equivalent to
\begin{equation}\label{eq:k}
\mathrm{Pr}(S_f^2(w_0)> \frac{\epsilon^2}{4})\geq 1-\delta.
\end{equation}
\end{Proof}
\subsection{Generalization to a multi-output Boolean function}
Given a multi-output Boolean function $F:\{0,\,1\}^n\rightarrow\{0,\,1\}^m$, where $m>0$ is an integer. If we can realize a quantum oracle $U_{b\cdot F}$ access to $b\cdot F$ for every $b\in F_2^m$, then after query $O(\frac{2^m\log\frac{1}{\delta}}{\epsilon^4})$ times, we will find larger coefficients of every $b\cdot F$. That is to say, the coefficients satisfying \eqref{eq:e} with $b\cdot F$ substituting for $f$.

In fact, we do the following procedure Instead of applying $U_{b\cdot F}$ directly in Deutsch--Jozsa algorithm. Before giving the algorithm, let us see the inner product operator $U_{IP}$ induced via Toffoli gates \cite{BBC95}
\begin{equation}\label{eq:la}
U_{IP}|x\rangle^{\otimes n}|y\rangle^{\otimes n}|-\rangle
=(-1)^{x\cdot y}|x\rangle^{\otimes n}|y\rangle^{\otimes n}|-\rangle,
\end{equation}
which is appeared in \cite{CGXL19} with concrete circuit.

\textbf{Algorithm: Quantum Walsh transform of $b\cdot F$}

1. The initial state is
\begin{equation}\label{eq:l}
|\psi_0\rangle=|0\rangle^{\otimes n}
|0\rangle^{\otimes m}|b\rangle^{\otimes m}|1\rangle.
\end{equation}

2. Apply Hadamard transformation to the first and fourth registers producing
\begin{equation}\label{eq:m}
|\psi_1\rangle=\sum_{x\in F^n_2}
\frac{1}{\sqrt{2^n}}|x\rangle^{\otimes n}|0\rangle^{\otimes m}|b\rangle^{\otimes m}
\frac{|0\rangle-|1\rangle}{\sqrt{2}}.
\end{equation}

3. Apply the $F$-controlled-NOT gate on the first and second registers to get
\begin{equation}\label{eq:n}
|\psi_2\rangle=\sum_{x\in F^n_2}
\frac{1}{\sqrt{2^n}}|x\rangle^{\otimes n}|F(x)\rangle^{\otimes m}|b\rangle^{\otimes m}
\frac{|0\rangle-|1\rangle}{\sqrt{2}}.
\end{equation}

4. Apply the inner product operator $U_{IP}$ to the second, third and fourth registers to get
\begin{equation}\label{eq:o}
|\psi_3\rangle=\sum_{x\in F^n_2}(-1)^{b\cdot F(x)}
\frac{1}{\sqrt{2^n}}|x\rangle^{\otimes n}|F(x)\rangle^{\otimes m}|b\rangle^{\otimes m}
\frac{|0\rangle-|1\rangle}{\sqrt{2}}.
\end{equation}

5. Apply Hadamard transform to the first register to obtain
\begin{equation}\label{eq:p}
|\psi_4\rangle=\sum_{a\in F^n_2}\frac{1}{2^n}
\sum_{x\in F^n_2}(-1)^{a\cdot x+b\cdot F(x)}
{|a\rangle}^{\otimes n}|F(x)\rangle^{\otimes m}|b\rangle^{\otimes m}
\frac{|0\rangle-|1\rangle}{\sqrt{2}}.
\end{equation}

6. Measure the first register in the computational basis.

Next, we use the above quantum walsh transform for every $0\neq b\in F_2^m$ to get the following algorithm.

\begin{bframe}
\textbf{Algorithm 2}

For any $ 0<\epsilon\leq 1$, $0<\delta<1$,
let $l=\frac{8\log\frac{1}{\delta}}{\epsilon^4}$,
$s=\frac{\epsilon^2l}{2}=\frac{4\log\frac{1}{\delta}}{\epsilon^2}$.
$H=\emptyset$, $L=\emptyset$ are two sets, where $\emptyset$ is the empty set.

\;for all $b\in[\,1,\,2^m-1\,]$ do

\;\quad for all $k\in [\,1,\,l\,]$ do

\;\quad\quad Run the quantum walsh transform of $b\cdot F$ to get $n$-bit vector $a$

\;\quad\quad If $a\in H$, then

\;\quad\quad\quad $i_a=i_a+1$

\;\quad\quad else

\;\quad\quad\quad update $H:=H\bigcup\{a\}$

\;\quad\quad end if

\;\quad\quad If $i_a\geq s$, then

\;\quad\quad\quad update $L:=L\bigcup\{(a,b)\}$

\;\quad\quad end if

\;\quad end for

\; end for

\;return $L$.

\end{bframe}

Through an analog analysis, we can prove the following result.

\begin{Theo}\label{th:b}
Given a vectorial Boolean function $F:\{0,\,1\}^n\rightarrow\{0,\,1\}^m$ and a threshold $0<\epsilon\leq 1$, running the above Algorithm 2
$O(\frac{2^{m}}{\epsilon^4}\log \frac{1}{\delta})$ times output a list $L=\{(a,b)_1,\ldots,(a,b)_l\}$ such that
\begin{equation}\label{eq:q}
\begin{cases}
|S_{b\cdot F}(a)|\geq \epsilon\Rightarrow (a,b)\in L,\\
(a,b)\in L\Rightarrow |S_{b\cdot F}(a)|\geq \epsilon/2.
\end{cases}
\end{equation}
with probability at least $1-\delta$.
\end{Theo}

\section{Conclusion}
This paper designs a quantum algorithm to obtain some large Walsh coefficients of a Boolean function with $O(\frac{\log\frac{1}{\delta}}{\epsilon^4})$ quantum queries, while classical algorithm in \cite{O14} uses $O(n\frac{\log\frac{1}{\delta}}{\epsilon^6})$ queries with the same probability and accuracy. Then, we generalize the quantum algorithm to apply to multi-output Boolean functions with a query complexity $O(2^m\frac{\log\frac{1}{\delta}}{\epsilon^4})$, compared with the quantum algorithms in \cite{CGXL19} with query complexity $O(\frac{2^{m+9}n\pi}{\epsilon^4}\log\frac{2^{m+3}n}{\delta\epsilon^2})$ and
$O(\frac{2^{m+5+n/2}}{\epsilon^3}\log\frac{2^{m+5}n}{\delta\epsilon^2})$ separately.

\subsubsection*{Acknowledgments.}

This work was supported by the Science and Technology Project of Henan Province (China) under Grant No.162102210103.

\end{document}